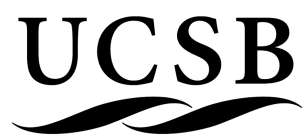

# Feature Selection and Model Comparison on Microsoft Learning-to-Rank Data Sets

Sen LEI, Xinzhi HAN

*Author: Sen LEI, Xinzhi HAN*

University of California, Santa Barbara

Xinzhi Han, Sen Lei

Feature Selection and Model Comparison on Microsoft Learning-to-Rank Data Sets

Abstract

With the rapid advance of the Internet, search engines (e.g., Google, Bing, Yahoo!) are used by billions of users for each day. The main function of a search engine is to locate the most relevant webpages corresponding to what the user requests. This report focuses on the core problem of information retrieval: how to learn the relevance between a document (very often webpage) and a query given by user.

Our analysis consists of two parts: 1) we use standard statistical methods to select important features among 137 candidates given by information retrieval researchers from Microsoft. We find that not all the features are useful, and give interpretations on the top-selected features; 2) we give baselines on prediction over the real-world dataset MSLR-WEB by using various learning algorithms. We find that models of boosting trees, random forest in general achieve the best performance of prediction. This agrees with the mainstream opinion in information retrieval community that tree-based algorithms outperform the other candidates for this problem.

*Author: Sen LEI, Xinzhi HAN*

# Contents

*Author: Sen LEI, Xinzhi HAN*

**The DATA SET that we use is available on [Microsoft Learning to Rank Datasets](#).**

**The CODE we wrote is available on [GitHub](#).**

Author: Sen LEI, Xinzhi HAN

# Chapter 1

# Introduction

In this paper we present our experiment results on Microsoft Learning to Rank dataset MSLR-WEB [20]. Our contributions include:

- Select important features for learning algorithms among the 136 features given by Microsoft.

- Given baseline evaluation results and compare the performances among several machine learning models.

To our knowledge we are the first to select features on the dataset MSLR-WEB, and we give baseline results on the models that are not covered by existing works. To make sure our results are reproducible, all of our scripts are available online, and detailed experiment procedures are given.

## 1.1 Background and Related Works

Search engines, or information retrieval at large, plays an important role in modern Internet. Given any query from the user, an ideal search engine should match the related web pages, and rank them based on relevance with the query. Since very often the user only takes a look at the top-ranked webs, it is crucial to locate the most relevant web pages. Hence it is interesting to learn the relevance between the query and web page by data mining algorithms. A line of works called *Le*arning *to R*ank (LetoR) [13, 27, 24, 4] focused on this learning problem, and several algorithms have been proposed, e.g., RankSVM [13], RankBoost [24], AdaRank [27], LambdaMART [4], etc. Some of these algorithms are applied in commercial search engines such as Google, Bing and Yahoo [16].

To better research LetoR, Microsoft Research and Yahoo have provided large scale datasets: LetoR 3.0 [22], LetoR 4.0 [21], MSLR-WEB [20] Yahoo Challenge [6]. Unlike Yahoo Challenge dataset, Microsoft Research has given descriptions on how to generate the features of all their datasets. Hence Microsoft datasets [22, 21, 20] are more useful in research. However, there are two challenges with the datasets MSLR-WEB:

- **Insufficient baselines reported**. For LetoR 3.0 and LetoR 4.0, baseline results are posted on the website [22, 21], and they have extensive research works presenting experiment results on them [19, 28, 6, 18]. For MSLR-WEB, the authors did not give baselines. To our knowledge there are only two existing works reporting experiment results on MSLR-WEB [2, 25]. However in [2, 25] only limited models and evaluation metrics are reported. Some competitive learning algorithms, e.g., Generalized Linear Model, Logistic Regression, SVM,

Random Forest (very promising based on a recent report [12]), Boosting Regression Tree, are not reported. In this paper we fill this void, and report both precision and Normalized Discounted Cumulative Gain (NDCG) following the dataset authors' baselines [22]. Hence our results should be comparable with the existing works [2, 25].

- **Too many features**. MSLR-WEB has 136 features. Note that each of LetoR datasets has only less than 50 features. It is interesting to investigate the significance of each of the 136 features. This paper will try a feature selection. Another issue is that some features in MSLR-WEB are developed and privately owned only by Microsoft (e.g., the features on user click data, boolean model and language model). This will make the dataset not fully reproducible. By feature selection, we can evaluate the importance of these private features. If some of them do not significantly influence the learning performance, we can simply discard them in the future research, when we need to generate features for new query and document datasets such as TREC Robust05 [26], TREC Enterprise05 [8], Clueweb09 [5], etc.

## 1.2 Dataset Description

Tables 1.1, 1.2 and 1.3 are an adaptation of the feature list given by MSLR-WEB website [20]. All the features are numeric.

The original datasets have the matched document for at most 30,000 queries, separated in 5 folds for five-fold cross validation. However in this report we only use the Fold 1 data, in order to reduce the training time into an appropriate scale (e.g., only training a RankNet or Coordinate Ascend model on Fold 1 of MSLR-WEB30K without cross validation over the five folds took more than 2 days). The results in this report only serve as baselines for future comparison use.

## 1.3 Our Contributions

There are two main contributions:

- We use Lasso on logistic regression, Lasso on ordinal regression, SVM, random forest, and generalized boosting models to select significant features among the 136 features from the dataset MSLR-WEB. We list the top-selected features and give interpretations on their importance.

- We report metrics Precision and NDCG for the models including Lasso on logistic regression, Random Forest, Generalized Boosted Regression, SVM, and Continuation Ratio model. We also give baselines for the other state-of-art LetoR algorithms.

*Author: Sen LEI, Xinzhi HAN*

Table 1.1: Description of the MSLR-WEB dataset (Part 1).

| Feature | No. | Description |
| --- | --- | --- |
| covered query term number | 1 - 5 | How many terms in the user query are covered by the text. The text can be body, anchor, title, url and whole document (for features 1 - 5 respectively, similarly below). |
| covered query term ratio | 6 - 10 | Covered query term number divided by the number of query terms. |
| stream length | 11 - 15 | Text length. |
| IDF (inverse document frequency) | 16 - 20 | 1 divided by the number of documents containing the query terms. |
| sum of term frequency | 21 - 25 | Sum of counts of each query term in the document. |
| min of term frequency | 26 - 30 | Minimum of counts of each query term in the document. |
| max of term frequency | 31 - 35 | Maximum of counts of each query term in the document. |
| mean of term frequency | 36 - 40 | Average of counts of each query term in the document. |
| variance of term frequency | 41 - 45 | Variance of counts of each query term in the document. |

Table 1.2: Description of the MSLR-WEB dataset (Part 2).

| Feature | No. | Description |
|---|---|---|
| normalized sum of stream length | 46 - 50 | Sum of term counts divided by text length. |
| normalized min of stream length | 51 - 55 | Minimum of term counts divided by text length. |
| normalized max of stream length | 56 - 60 | Maximum of term counts divided by text length. |
| normalized mean of stream length | 61 - 65 | Average of term counts divided by text length. |
| normalized variance of stream length | 66 - 70 | Variance of term counts divided by text length. |
| sum of tf*idf | 71 - 75 | Sum of the product between term count and IDF for each query term |
| min of tf*idf | 76 - 80 | Minimum of the product between term count and IDF for each query term |
| max of tf*idf | 81 - 85 | Maximum of the product between term count and IDF for each query term |
| mean of tf*idf | 86 - 90 | Average of the product between term count and IDF for each query term |
| variance of tf*idf | 91 - 95 | Variance of the product between term count and IDF for each query term |
| boolean model | 96 - 100 | **Unclear. Privately owned by Microsoft.** |
| vector space model | 101 - 105 | dot product between the vectors representing the query and the document. **The vectors are privately owned by Microsoft.** |
| BM25 | 106 - 110 | Okapi BM25 |
| LMIR.ABS | 111 - 115 | Language model approach for information retrieval (IR) with absolute discounting smoothing [17] |
| LMIR.DIR | 116 - 120 | Language model approach for IR with Bayesian smoothing using Dirichlet priors [3] |
| LMIR.JM | 121 - 125 | Language model approach for IR with Jelinek-Mercer smoothing [17] |

*Author: Sen LEI, Xinzhi HAN*

Table 1.3: Description of the MSLR-WEB dataset (Part 3).

| Feature | No. | Description |
| --- | --- | --- |
| number of slashes in URL | 126 | e.g., "ucsb.edu/pstate/people" has 2 slashes. |
| length of url | 127 | The number of characters in the URL. |
| Inlink number | 128 | The number of web pages that cite this web page. |
| Outlink number | 129 | How many web pages this web cites. |
| PageRank | 130 | Evaluates the centrality of this web page based on web links over the Internet. This gives the success of Google. |
| SiteRank | 131 | Site level PageRank. E.g., "ucsb.edu/pstat" and "ucsb.edu/math" share the same SiteRank. |
| QualityScore | 132 | The quality score of a web page. The score is outputted by a web page quality classifier. **Privately owned by Microsoft.** |
| QualityScore2 | 133 | The quality score of a web page. The score is outputted by a web page quality classifier, which measures the badness of a web page. **Privately owned by Microsoft.** |
| Query-url click count | 134 | The click count of a query-url pair at a search engine in a period. **Collected and privately owned by Microsoft.** |
| url click count | 135 | The click count of a url aggregated from user browsing data in a period. **Collected and privately owned by Microsoft.** |
| url dwell time | 136 | The average dwell time of a url aggregated from user browsing data in a period. **Collected and privately owned by Microsoft.** |

*Author: Sen LEI, Xinzhi HAN*

# Chapter 2

# Data Processing and Models

## 2.1 Data Processing

Speaking of our training data which contains 137 numerical variables based on 10000 observations, thus a 137 by 10000 data frame, due to extremely expansive computation, we randomly sampled those observations out of over 220000 original observations included in, as previously referred, Fold 1 of MLSR-WEB30K. Among all the 137 variables, the first one is the response variable named `rel`, standing for web page relevance. The rest 136 variables, denoted as `X1`, `X2`, $\cdots$, `X136`, are independent variables, out of which the first 125 variables consist of 5 perspectives that are body, anchor, title, url, and whole document within each 25 larger features. Some of these features are public and well-known, such as `query term number`, which is the number of terms in a users query covered by text; stream length, describing the text length; different aggregations of term frequency, that are aggregations on counts of each query term in a document; `tf-idf`, short for term frequencyinverse document frequency, etc. Apart from that, there are also 11 independent variables not of above 5 perspectives. Despite some of them being privately designed by Microsoft and remained unclear, others are not mysterious, e.g., `PageRank`, a ranking criterion used by Google search engine; `length of url`, namely, the number of characters in a url; Outlink number, a web page quality about the number of other web pages cited by this web page. Our test data, of the same schema as the training data, includes about 240000 observations collected in 1961 different files. We will conduct predictions using these test files. Our goal aims at comparing several trained models based on predicted measurements of accuracy.

## 2.2 Binary Classification

Take a closer look at the response variable `rel`, we notice that even though it has 5 relevance levels ranged from 0 to 4, observations at relevance level 3 are insufficient and are only about 1.5% of total 10000 observations. What is worse, those at level 4 are even less that 1%, which may lead to potential over-fitting, hence unpersuasive result. In that case, we realize it might be better to combine levels with fewer observations together. We then treat levels of $0, 1$ as our new level 0, and levels of $2, 3, 4$ as the new level 1. By doing that, our task becomes a binary classification which allows more tools for us to deal with.

### 2.2.1 LASSO in Binomial Regression

In our intuition, logistic regression does a good job in binary classification, so we first decide it to be our first model. The interesting thing is that after fitting a logistic model, singularity occurs, indicating some of the variables have perfect collinearity, that is to say, some of the variables are exact linear combinations of others. However, in this project we are more interested in Support Vector Machine on binary classification the predictive ability of our model rather than individual coefficients. Aliased variables which do not contribute to the model will not affect our model accuracy. Carrying on from that, still, we think 136 features are way too inefficient and expansive to do computations, and we do not expect each of these features to be relevant for later prediction, therefore the motivation here is to find a penalized model and apply what is called shrinkage on those features to tell us what variables are important and thus should be kept, what contribute little to the model and therefore can be thrown away. Lasso, short for least absolute shrinkage and selection operator, trying to minimize the objective function

$$\sum_{i=1}^{n}\left(y_i - \beta_0 - \sum_{j=1}^{p} x_{ij}\beta_j\right)^2 + \lambda \sum_{j=1}^{p} |\beta_j|$$

is exactly what we are looking for. It shrinks the estimated coefficients to actual 0s as the increase of the tuning parameter $\lambda$, but nonetheless we have to be aware when to stop penalizing and keep the rest non-zero coefficients. The idea behind that is the bias variance trade-off. To achieve that, we use R function `cv.glmnet` [10] that performs a cross-validation on the training set to get the best $\lambda$ – lambda.1se within the result from cv.glmnet, and then use this to select the best subset of the features which contains 12 variables. Later, instead of predicting the measurements of accuracy using these 12 variables, we reset tuning parameter from lambda.1se to lambda.min to include more features in order to increase prediction accuracy but without worrying about risks from collinearity. We used test sets to conduct predictions on the response scale and obtained 1961 probability vectors of length n with each specifying the probability that each of the n observations is assigned to label 1. We will use these vectors to calculate the measurements of accuracy Precision and NDCG. We will be talking about these measurements as well as the selected features in detail later in Chapter 3.

### 2.2.2 Support Vector Machine on binary classification

Support vector machine is a reasonable method for binary classification problems, it solves an optimization problem looks like [7]

$$\max_{\beta_1,\cdots,\beta_p} M$$

$$s.t. \begin{cases} \sum_{j=1}^{p} \beta_j^2 = 1 \\ y^{(i)}\left(\beta_0 + \beta_1 x_1^{(i)} + \cdots + \beta_p x_p^{(i)}\right) \geq M(1 - \epsilon^{(i)}) \\ \epsilon^{(i)}) \geq 0, \sum_{i=1}^{n} \epsilon^{(i)}) \leq C \end{cases}$$

where $\beta_j$ are coefficients; $M$ is the maximum margin distance; $\epsilon(i) > 0$ are slack variables, which means the number of observations allowed to be mis-classified in Non-separable case; $C$ is the "budget": a tuning parameter which controls the amount of slack. We are curious whether it is a good model for our predictions, or at least better than the previous binary case Lasso model. Due to time constraints, we are not able to compare the performance of each kernel through some parameter optimization experiments. We will stick to the linear kernel, which only involves a cost parameter (an inverse version of the budget parameter $C$) to optimize, that allows us to focus more on our classification models. For the purpose of saving time, to tune the cost parameter, instead of using the whole data, we use a random sub-sample of size 500. We also fix the tuning range to be $(0.001, 0.01, 0.1, 1, 5, 10, 100)$. After tuning, we obtained the best `cost = 0.001` over that range. Knowing that later we need to calculate the model precision which involves sorting probabilities, we would rather want an explicit probability for the class labels; therefore, we create 50 bootstrap replicates of the training data to estimate class probabilities within each test document. After that, we acquired 1961 number of $n$ by 2 probability matrices, within each of which 2 columns names are specified as class label `"0"` and `"1"` and n rows names are indices of n observations in that particular file. To have an idea on important features, we apply R function `rfe` within package caret [14], which is about recursive feature elimination algorithm, to select the best 10 features out of 136 using method `"svmLinear"` and 20 folds cross validation.

## 2.3 Multiclass Classification

One problem with binary classification is that we lose certain information among labels; how different is different? Is the difference between level 3 and 4 the same as that between level 1 and 4? Take this into consideration, we are going to build some multiclass classification models and to compare the accuracy with binary classification models.

### 2.3.1 Continuation Ratio Model

Considering the fact that the response variable is ordinal, and there are too many covariates, we decided to try penalized Continuation Ratio Model.

A variety of statistical modeling procedures, namely, proportional odds, adjacent category, stereotype logit, and continuation ratio models can be used to predict an ordinal response. In this paper, we focus attention to the continuation ratio model because its likelihood can be easily re-expressed such that existing software can be readily adapted and used for model fitting [1].

**Statistical Background**

Suppose for each observation, $i = 1, ..., n$, the response $Y_i$ belongs to one ordinal class $k = 1, \cdots, K$ and $\mathbf{x}_i$ represents a $p$-length vector of covariates. The backward formulation of the continuation ratio models the logit as

$$logit\left[\mathbb{P}(Y = k | Y \leq k, X = x)\right] = \alpha_k + \beta_k^T \mathbf{x}$$

For high dimensional covariate spaces, the best subset procedure is computationally prohibitive. However, penalized methods, places a penalty on a function of the coefficient estimates, permitting a model fit even for high-dimensional data [23].

A generalization of these penalized models can be expressed as,

$$\tilde{\beta} = \arg\min_{\beta} \left[ \sum_{i=1}^{n} \left( y_i - \beta_0 - \sum_{j=1}^{p} x_{ij}\beta_j \right)^2 + \lambda \sum_{j=1}^{p} |\beta_j|^q \right], \quad q \geq 0$$

When $q = 1$ we have the $L_1$ penalized model, when $q = 2$ we have ridge regression. Values of $q \in (1, 2)$ provide a compromise between the $L_1$ and ridge penalized models. Because when $q > 1$ coefficients are no longer set exactly equal to 0, the elastic net penalty was introduced [1]

$$\lambda \sum_{j=1}^{p} \left( \alpha \beta_j^2 + (1 - \alpha)|\beta_j| \right)$$

*Author: Sen LEI, Xinzhi HAN*

**Model Building**

We separate the original data randomly into a 10000-*obs* training set and a 2000-*obs* test set.

We use the package `glmnetcr` in R to fit the model.

Figure 2.1 can be used to identify a more parsimonious model having a BIC close to the minimum BIC; finally we chose the model when `step = 32`.

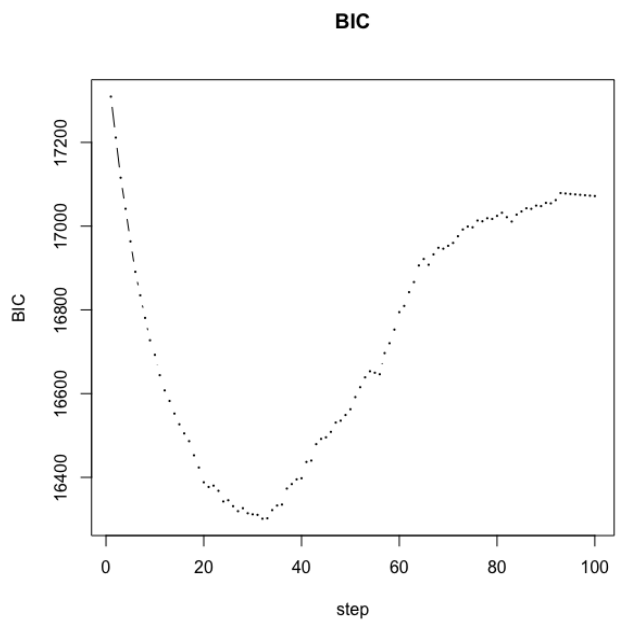

Figure 2.1: Plot of Bayesian Information Criteria across the regularization path for the fitted `glmnetcr` object using the training subset of lector data set.

Then, I use the remaining 2000 test set to make the prediction. The evaluation methods is discussed in Chapter 3, and evaluation results are shown in Table 3.6 and 3.7.

### 2.3.2 Random Forest

The motivation of using Random Forest method is unlike ordinary bagged decision tree model, it chooses split variable from a random subset of the predictors, in which case collinearity issues will not be caused by highly correlated variables we presumed to exist. From error reduction perspective, it achieves bias variance trade-off by reducing variances of large complex models. First of all, we use R function `rfcv` in the package `randomForest` to try sequentially variable importance pruning via a nested cross-validation procedure. We set the fraction of variables to remove at each iteration to be 0.7.

Table 2.1: Part of Cross-validation error in Random Forest for feature selection

| 136 | 95 | 67 | 47 | 33 | 23 | 16 | 11 | 8 | 5 |
| --- | --- | --- | --- | --- | --- | --- | --- | --- | --- |
| 0.45075 | 0.45063 | 0.45738 | 0.45475 | 0.46375 | 0.46600 | 0.47288 | 0.47325 | 0.47950 | 0.48925 |

From Table 2.1, we notice that the CV error increases as the number of predictors are reduced, and the error difference between using 136 features and 95 features is very low, which suggests the 136-feature model is as good as the 95-feature model, so we decide to use original 136 features to fit our random forest model. In R function, we set argument `ntree` to a reasonable size 501, importance to be true since we want our model to assess the importance of features, and keep other arguments their default settings, where in our case, the number of variables randomly sampled at each split is $\sqrt{136} \approx 12$. After fitting the model, we predicted using our test sets on probability scale and get 1961 $n$ by 5 probability matrices. 5 columns names in each of these matrices are specified as class label "0", "1", "2", "3", and "4"; n rows names are indices of n observations in that particular file.

### 2.3.3 Generalized Boosted Regression Modeling

On the other hand, not like random forest, boosting achieves bias variance trade-off by reducing biases of low-variance models, which inspires us to see how this model works. The brief idea behind boosting is that it fits trees multiple times sequentially, and uses information (fraction of mis-classifications) from previous grown trees as weak learners to update classifier weights after each iteration, then calculates a weighted average of weak learners' classifications as the final prediction [11]. Before fitting a boosting model with designed parameters, we need to figure out what a good value is for each of the parameters. Here, we are interested in finding out appropriate values for parameters `n.trees` and `interaction.depth` within R function `gbm`. Fortunately, there is a package in R called caret that does this job. This package is for classification and regression training, where we decide appropriate ranges for parameters we want to train, and the function `caret::train` [14] will tune the best parameters (with smallest cross validation error) for us. We predetermined our tree size to be from 600 to 1500 with increment 125, and the maximum depth of variable interactions to be 2, 4, and 6. We set the shrinkage, AKA learning rate, to 0.001, because usually the smaller the rate is, the more precise the model will be. We sampled 1000 observations for the tuning process using 20 folds cross validation and obtained desired parameters as `n.trees = 1250` and `interaction.depth = 4`. Using these parameters to fit our boosting model after setting distribution family to `"multinomial"`, a summary of this fitted model gives us ranked importance of each feature presented by a plot as well as a table.

*Author: Sen LEI, Xinzhi HAN*

# Chapter 3

# Evaluation Results

We here present our experiment setup and baseline results. To make sure our results are reproducible, we make all of our experiment scripts available [15].

## 3.1 Setup and Measurements

We implemented our algorithms by using R, and we also used the state-of-art algorithms given by Ranklib [9]. We leave all of the baselines generated by Ranklib to the Appendix.

We use Precision and NDCG as the measurement of accuracy, as suggested by the authors [20]. These two measurements are widely used in the existing works [22, 20].

- Given any query, precision $P@k$ is defined as $R_k/k$, where $R_k$ the number of truly relevant documents among the top $k$ documents selected by the learning algorithm. That is, given any query, the learning algorithm should give a ranking of documents base on the predicted relevance, and we want to see how many documents of the top $k$ are really relevance by the ground truth. Note that the ground truth relevance in MLSR-WEB has five levels: 0, 1, 2, 3, and 4. As suggested by the dataset authors [20], we regard 0 and 1 as irrelevance (i.e., 0) and regard 2, 3, 4 as relevance (i.e., 1) when we evaluate precision.

- Given any query, NDCG is defined as

$$NDCG@k = \frac{DCG@k}{IDCG@k},$$

where DCG is defined as

$$DCG@k = \sum_{i=1}^{k} \frac{(2^{rel_i} - 1)}{\log_2(i+1)},$$

where $rel_i$ denotes the true relevance of the $i$-th ranked document as suggested by the learning algorithm, and IDCG is defined as

$$IDCG@k = \sum_{i=1}^{k} \frac{(2^{ideal_i} - 1)}{\log_2(i+1)},$$

where $ideal_i$ denotes the true relevance of the $i$-th rankled document if we rank the matched documents by their ground truth relevance.

Note that NDCG has several variants. Here we use the version from MLSR-WEB evaluation script [20].

*Author: Sen LEI, Xinzhi HAN*

Here we argue why we use information retrieval measurements such as precision and NDCG, rather than traditional statistical errors such as Mean Squared Error (MSE) or Area Under Curve (AUC) of ROC. There are basically two reasons.

1. The relevance scores are only qualitative and very subjective. Note that the relevance levels from 0 to 4 are given by human experts. Thus these relevance levels by no means can be quantitatively accurate, i.e., they only *roughly* represent how people feel the relevance between a query and a document. For example, relevance level 4 does not imply its relevance is twice of relevance level 2. However most traditional statistical measurements assume the targets are quantitatively accurate.

2. Only the top ranked documents are considered for evaluation. This is typically how people use the search engine: send some query to the search engine, and only take a look at the *very* first ranked documents. Thus an appropriate measurement for evaluation should pay overwhelming weights on the top-ranked documents, like Precision and NDCG. However both AUC and MSE give *equal* weights to each document in the test dataset.

The dataset MSLR-WEB10 has 10,000 queries, MSLR-WEB30K has 30,000 queries. On average, for each query, there are 100 - 200 matched documents that have relevance levels evaluated by human experts. We average the measurements over all the involved queries.

We run our experiments on UCSB Center of Scientific Computing, Cluster Knot's 93-th node, which is a DL580 node with 4 Intel X7550, eight core processors and 1TB of RAM.

Now refer Precision and NDCG to each of our models:

## 3.2 Evaluation results of each model

### 3.2.1 LASSO in Binomial Regression

As we stated above, setting tuning parameter $\lambda$ to `lambda.1se` which is acquired in cross validation Lasso gives us 12 non-zero coefficient variables listed below, thus, we consider these 12 features to be the most important ones. Notice that there is no order on the importance of these 12 features.

Table 3.1: Selected features under Binomial Lasso Regression

| Index | Feature |
|---|---|
| X28 | min of term frequency in title |
| X30 | min of term frequency in whole document |
| X64 | mean of stream length normalized term frequency in URL |
| X65 | mean of stream length normalized term frequency in whole document |
| X98 | boolean model in title |
| X108 | BM 25 in title |
| X109 | BM 25 in URL |
| X123 | LMIR.JM in title |
| X126 | Number of slash in URL |
| X127 | Length of URL |
| X129 | Outlink number |
| X130 | PageRank |

From the table, we find that entire selected features are related with `term frequency`, `stream length`, `boolean model`, BM25, LMIR.JM, URL, and `PageRank`.

To calculate the Precision, we already have 1961 probability vectors of length n, where n depends on the number of observations within each test file. For each of these vectors, we sort the observations in a decreasing order according to their corresponding probabilities and fetch the first k indices. Back to the original test data, we look at the class labels of observations associated with these k indices and calculate the potion of observations assigned to label 1. Roughly speaking, we would like to know the actual percentage of class label 1 among selected k observations given that those k observations are predicted to be in label 1. Finally, we take an average of all the percentages of all the test files to be our Precision. The reason of choosing a relatively small k number of observations is because in real life, a client is usually interested in the first k query results and disregards the rest. The process of calculating the NDCG largely based on the above given formula. The results of both measurements are presented in the Tables 3.6 and 3.7.

### 3.2.2 Support Vector Machine

Selected features are listed in the table below (Order matters).

Table 3.2: Selected features under Support Vector Machine

| Index | Feature |
|---|---|
| X55 | min of stream length normalized term frequency in whole document |
| X78 | min of tf*idf in title |
| X80 | min of tf*idf in whole document |
| X65 | mean of stream length normalized term frequency in whole document |
| X50 | sum of stream length normalized term frequency in whole document |
| X51 | min of stream length normalized term frequency in body |
| X76 | min of tf*idf in body |
| X88 | mean of tf*idf in title |
| X60 | max of stream length normalized term frequency in whole document |
| X30 | min of term frequency in whole document |

Features selected in SVM model are mainly within the scope of `stream length`, `term frequency`, and `tf-idf`.

The process of calculating the Precision and NDCG are similar to what we did in Lasso binomial regression model, and the result is listed in Table 3.6 and 3.7.

### 3.2.3 Continuation Ratio Model

As we stated above, using BIC, R gives us 26 non-zero coefficient variables listed in Table 3.3, thus, we consider these 12 features to be the most important ones. Notice that there is no order on the importance of these 26 features.

From the table, one can find that features selected under continuation ratio model are within categories `query term`, `stream length`, `term frequency`, BM25, LMIR, `PageRank`, and URL.

Precision and NDCG can be calculated similarly to those in random forest model, and results are given in Table 3.6 and 3.7.

*Author: Sen LEI, Xinzhi HAN*

Table 3.3: Selected features under Continuation Ratio Model

| Index | Feature |
|---|---|
| X3 | covered query term number in title |
| X4 | covered query term number in url |
| X6 | covered query term ratio in body |
| X11 | stream length in body |
| X15 | stream length in whole document |
| X22 | sum of term frequency in anchor |
| X26 | min of term frequency in body |
| X30 | min of term frequency in whole document |
| X41 | variance of term frequency in body |
| X45 | variance of term frequency in whole document |
| X49 | sum of stream length normalized term frequency in url |
| X70 | variance of stream length normalized term frequency in whole document |
| X72 | sum of tf*idf in anchor |
| X98 | boolean model in title |
| X107 | BM25 in anchor |
| X108 | BM25 in title |
| X109 | BM25 in url |
| X110 | BM25 in whole document |
| X115 | LMIR.ABS in whole document |
| X123 | LMIR.JM in title |
| X126 | Number of slash in URL |
| X129 | Outlink number |
| X130 | PageRank |
| X133 | QualityScore2 |
| X134 | Query-url click count |
| X136 | url dwell time |

### 3.2.4 Random Forest

Random forest showed us the most important 10 features in an importance plot, shown in figure 3.1.

Taking both plots into consideration, we made Table 3.4 containing selected features under random forest model (Order matters).

Selected features under random forest model are mostly related to `PageRank`, `stream length`, `SiteRank`, `URL`, and `LMIR`.

Figuring out Precision for multi-class classification involves a concept called "expected rank". Each observation has an expected rank, which can be calculated by the summing up the product of each class label and its corresponding probability. Then again, we sort the observations in a decreasing order according to their expected ranks and fetch the first $k$ indices. Then we compare and find the proportion of actual class label 1 in each of our test set and take the average. Notice here after calculating and sorting the expected ranks, we acquire the proportion by converting labels in test files to `"0"` and `"1"` using the same rule we did in binary classification.

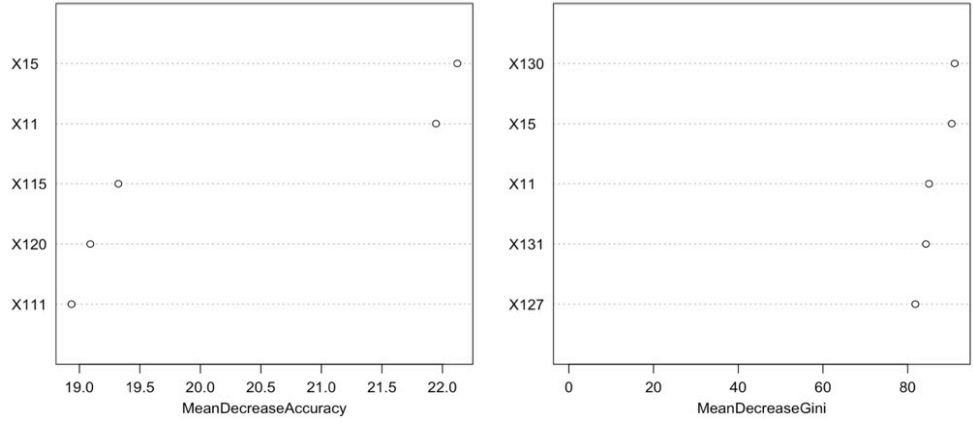

Figure 3.1: Importance plot under random forest model

Table 3.4: Selected features under Random forest model

| Index | Feature |
|---|---|
| X130 | PageRank |
| X15 | stream length in whole document |
| X11 | stream length in body |
| X131 | SiteRank |
| X127 | Length of URL |
| X115 | LMIR.ABS in whole document |
| X120 | LMIR.DIR in whole document |
| X111 | LMIR.ABS in body |

### 3.2.5 Generalized Boosted Regression Modeling

We summarize the boosting model and directly get the feature importance as shown in Table 3.5.

Table 3.5: Selected features under GBM

| Index | Feature |
|---|---|
| X55 | min of stream length normalized term frequency in whole document |
| X88 | mean of tf*idf in title |
| X53 | mean of stream length normalized term frequency in title |
| X15 | stream length in whole document |
| X51 | min of stream length normalized term frequency in body |
| X123 | LMIR.JM in title |
| X115 | LMIR.ABS in whole document |
| X103 | vector space model in title |
| X134 | Query-url click count |
| X11 | stream length in body |

Selected features under boosting model are around `stream length`, `tf-idf`, `LMIR`, `vector space`, `Query-url click`.

Precision and NDCG can be calculated similarly to those in random forest model, and results are given in Table 3.6 and 3.7.

## 3.3 Evaluation by R

Tables 3.6 and 3.7 give the results for our algorithms implemented by R. We observe that Random Forest has the best performance.

Here we give some insights and interpretations on our results:

- We use SVM model to do a binary classification, i.e., for each instance, the target can only be relevant (1) or irrelevant (0). We use the criteria above to transform the original 5-level relevance into the binary version when training. By this transformation we lose significant information on the strength and weakness of each relevance score. This should significantly decrease the prediction performance, especially when we are evaluating NDCG, which encourages to rank highly relevant documents at top places.

- All the methods we use here are only pointwise [12]. So far we have not given the baselines of pairwise and listwise methods (see in the Appendix). Our loss functions are not yet directly related to Precision or NDCG. Thus the results here should not be competitive with beyond-pointwise methods.

- Tree based models (Random Forest, Boosting Trees) outperform the others. This agrees with the new founding in LetoR [16, 4, 12].

Table 3.6: Precision results for MSLR-WEB10K Fold 1.

| Model | @1 | @2 | @3 | @4 | @5 | @6 | @7 | @8 | @9 | @10 |
|---|---|---|---|---|---|---|---|---|---|---|
| Lasso | 0.283 | 0.272 | 0.268 | 0.265 | 0.261 | 0.257 | 0.256 | 0.253 | 0.252 | 0.249 |
| Random Forest | 0.594 | 0.520 | 0.478 | 0.449 | 0.428 | 0.415 | 0.402 | 0.390 | 0.379 | 0.368 |
| SVM | 0.308 | 0.292 | 0.279 | 0.271 | 0.263 | 0.256 | 0.250 | 0.245 | 0.240 | 0.236 |
| Ordinal | 0.374 | 0.344 | 0.326 | 0.317 | 0.312 | 0.304 | 0.299 | 0.289 | 0.284 | 0.280 |
| Boosting Trees | 0.473 | 0.438 | 0.413 | 0.391 | 0.377 | 0.364 | 0.353 | 0.343 | 0.333 | 0.326 |

Table 3.7: NDCG results for MSLR-WEB10K Fold 1.

| Model | @1 | @2 | @3 | @4 | @5 | @6 | @7 | @8 | @9 | @10 |
|---|---|---|---|---|---|---|---|---|---|---|
| Lasso | 0.227 | 0.242 | 0.253 | 0.264 | 0.272 | 0.280 | 0.289 | 0.296 | 0.303 | 0.309 |
| Random Forest | 0.456 | 0.427 | 0.420 | 0.418 | 0.420 | 0.423 | 0.426 | 0.429 | 0.432 | 0.435 |
| SVM | 0.251 | 0.251 | 0.255 | 0.261 | 0.266 | 0.272 | 0.277 | 0.282 | 0.287 | 0.291 |
| Ordinal | 0.284 | 0.291 | 0.298 | 0.307 | 0.317 | 0.325 | 0.333 | 0.338 | 0.344 | 0.349 |
| Boosting Trees | 0.377 | 0.371 | 0.373 | 0.376 | 0.380 | 0.385 | 0.389 | 0.394 | 0.398 | 0.402 |

# Chapter 4

# Discussions

## 4.1 Interpretations on Important Features

We find the following feature categories are important in the 137 candidates of MSLR-WEB:

- **Term frequency based features**. Typical examples include TFIDF, BM25, cover ratio of the query, LMIR smoothing, etc. These features are significant in nature since people would like to see the pages containing the words requested. Furthermore, the term frequencies in body and title weigh more than other parts of the web page based on our results.

- **Link based features**. Typical such features include PageRank, SiteRank, In/Out link number, etc. The intuition is also clear: important web pages tend to be much more cited than the ordinary web pages (also called hubs). If one put all the web pages and the web links between them into one graph, naturally the important web page should have a central place. PageRank (the core of Google's search engine) proves to capture such centrality tightly. Note that these features are only document specific, i.e., it will not change given different queries. This implies a certain potion of a successful relevance evaluation should only focus on the document itself, regardless of the query.

- **Click based features**. Features 134, 135 and 136 are in this category. The intuition is that users tend to click the most interesting web pages and dwell for long enough time on the relevant pages. Unfortunately very often these features gathered from real users are private to the search engine companies.

- **URL lengths**. Important and popular web pages are likely to have short URLs, which are easy to remember. Also the number of slashes in a popular URL should not be too many.

- **Lengths of web pages or titles**. This is also known as stream length (features 11 - 15). The intuition behind is that longer pages are more likely to contain more useful information, which should attract users.

## 4.2 Interpretations on Unimportant Features

Our results also suggest that some features in MSLR-WEB are not very useful:

- **Variance features**. Typical such features include variance of TFIDF, (normalized) term frequencies. We agree that low variance of TF means the document is unlikely to have a huge bias on certain terms in the query. [1] However, the important of this tendency is difficult to argue and lacks enough experiment results to back up.

- **Inverse Document Frequency (IDF) based features**. We agree that the intuition behind this feature category is that the web page should contain novel information rather than copy from other sources. A low IDF implies the content in the web page is unique and can rarely be found somewhere else. However it is difficult to see the connection between this feature with relevance given a query. Note that these features are document specific like PageRank, but they seem not to be able to capture the web page quality well enough.

## 4.3 On the Baselines

We use R to generate the baselines of Precision and NDCG for the standard statistical models, and we give more baselines in the Appendix for the state-of-art LetoR algorithms. Since our models are somewhat classical, i.e., pointwise compared to the state-of-art which are mostly pairwise and listwise, in nature our baselines cannot outperform the ones given by MART, LambdaMART, etc. For Random Forest, we achieve the similar performance compared to the results given by Ranklib (see Appendix).

## 4.4 Limitations and Future Works

The biggest limitation in this report is that the training set size is too small (no more than 10k instances for each training). This is mostly because the models we used in R are not scalable, consuming too much time when the number of instances exceeds 10k. Hence we samples 10k instances from the entire dataset, which contains nearly 3.8 million instances. In the future we may resort to scalable models (e.g., the ones from Ranklib, tensorflow or scikit-learn) to better capture the whole dataset.

It is still very open for the feature selection problem. We cannot rule out the possibility that the selected features in this report are only subject to the MSLR-WEB dataset. For example the significance of variance and IDF is not clear in this dataset, but that does not imply these features are not useful in any other datasets. We will evaluate the feature importance for other LetoR datasets to make more solid conclusion.

---

[1] Here we give an example of the bias. Suppose the query is "international organized crime". A news webpage talking about *local* organized crime is irrelevant, even though it has high frequency on part of the query "organized crime".

*Author: Sen LEI, Xinzhi HAN*

# Chapter 5

# Acknowledgement

We are indebted for Shiyu Ji [1], who shared with us his suggestions and comments on this paper.

---

[1] shiyu@cs.ucsb.edu

Author: Sen LEI, Xinzhi HAN

# Appendix A

# More Baselines on the State-of-Art LetoR Algorithms

For a better comparison of our results, we give more baselines of the existing LetoR Algorithms generated by RankLib 2.5 [9]. We trained and tested the algorithms on Fold 1 only. We did not modify any model parameters (number of trees or leaves, bagging size, learning rate, etc.): only the ones default by RankLib are used.

- Tables A.1 and A.2 give the baselines of the LetoR algorithms on the dataset MSLR-WEB10K, which is sampled from MSLR-WEB30K [20]. Gradient Boosting Regression Trees (GBRT), Coordinate Ascent and Random Forests have the best performance.

- Tables A.3 and A.4 give the baselines of the LetoR algorithms on the dataset MSLR-WEB30K. GBRT has the best performance.

Note that since we did not do any cross validation or parameter tuning, these results only serve as baselines for future comparisons.

Table A.1: Precision results for MSLR-WEB10K Fold 1.

| Model | @1 | @2 | @3 | @4 | @5 | @6 | @7 | @8 | @9 | @10 |
|---|---|---|---|---|---|---|---|---|---|---|
| GBRT | **0.500** | **0.474** | **0.448** | **0.430** | **0.415** | **0.403** | 0.391 | 0.379 | 0.368 | 0.359 |
| RankNet | 0.116 | 0.131 | 0.133 | 0.133 | 0.132 | 0.132 | 0.135 | 0.136 | 0.136 | 0.137 |
| RankBoost | 0.373 | 0.340 | 0.320 | 0.307 | 0.298 | 0.289 | 0.282 | 0.276 | 0.271 | 0.268 |
| AdaRank | 0.414 | 0.385 | 0.365 | 0.348 | 0.333 | 0.323 | 0.312 | 0.301 | 0.290 | 0.283 |
| Coordinate Ascent | 0.411 | 0.390 | 0.387 | 0.387 | 0.390 | 0.393 | **0.397** | **0.399** | **0.402** | 0.405 |
| LambdaMART | 0.422 | 0.399 | 0.382 | 0.365 | 0.355 | 0.347 | 0.337 | 0.326 | 0.318 | 0.311 |
| ListNet | 0.099 | 0.108 | 0.114 | 0.119 | 0.124 | 0.124 | 0.126 | 0.128 | 0.130 | 0.130 |
| Random Forests | 0.367 | 0.365 | 0.365 | 0.372 | 0.377 | 0.384 | 0.391 | 0.397 | **0.402** | **0.407** |

Table A.2: NDCG results for MSLR-WEB10K Fold 1.

| Model | @1 | @2 | @3 | @4 | @5 | @6 | @7 | @8 | @9 | @10 |
|---|---|---|---|---|---|---|---|---|---|---|
| GBRT | 0.401 | 0.400 | 0.404 | **0.409** | **0.414** | **0.421** | **0.426** | **0.430** | **0.434** | **0.439** |
| RankNet | 0.116 | 0.130 | 0.138 | 0.145 | 0.151 | 0.158 | 0.164 | 0.170 | 0.175 | 0.180 |
| RankBoost | 0.277 | 0.284 | 0.290 | 0.297 | 0.306 | 0.312 | 0.318 | 0.324 | 0.330 | 0.335 |
| AdaRank | 0.340 | 0.333 | 0.334 | 0.335 | 0.337 | 0.340 | 0.344 | 0.347 | 0.349 | 0.353 |
| Coordinate Ascent | **0.502** | **0.447** | **0.415** | 0.394 | 0.375 | 0.359 | 0.348 | 0.337 | 0.328 | 0.319 |
| LambdaMART | 0.323 | 0.336 | 0.343 | 0.349 | 0.357 | 0.363 | 0.369 | 0.373 | 0.378 | 0.383 |
| ListNet | 0.109 | 0.119 | 0.129 | 0.137 | 0.145 | 0.151 | 0.158 | 0.164 | 0.169 | 0.175 |
| Random Forests | 0.452 | 0.419 | 0.396 | 0.387 | 0.373 | 0.365 | 0.355 | 0.349 | 0.341 | 0.335 |

Table A.3: Precision results for MSLR-WEB30K Fold 1.

| Model | @1 | @2 | @3 | @4 | @5 | @6 | @7 | @8 | @9 | @10 |
|---|---|---|---|---|---|---|---|---|---|---|
| GBRT | **0.543** | **0.496** | **0.467** | **0.447** | **0.429** | **0.414** | **0.400** | **0.388** | **0.378** | **0.368** |
| RankNet | 0.125 | 0.128 | 0.130 | 0.129 | 0.130 | 0.130 | 0.132 | 0.132 | 0.132 | 0.133 |
| RankBoost | 0.367 | 0.341 | 0.324 | 0.311 | 0.302 | 0.294 | 0.286 | 0.280 | 0.274 | 0.269 |
| AdaRank | 0.251 | 0.262 | 0.262 | 0.259 | 0.257 | 0.254 | 0.250 | 0.248 | 0.245 | 0.242 |
| Coordinate Ascent | 0.519 | 0.464 | 0.427 | 0.401 | 0.380 | 0.364 | 0.351 | 0.340 | 0.331 | 0.323 |
| LambdaMART | 0.449 | 0.423 | 0.402 | 0.382 | 0.370 | 0.356 | 0.345 | 0.337 | 0.329 | 0.322 |
| ListNet | 0.121 | 0.125 | 0.128 | 0.130 | 0.131 | 0.133 | 0.134 | 0.137 | 0.138 | 0.138 |
| Random Forests | 0.462 | 0.429 | 0.406 | 0.387 | 0.376 | 0.366 | 0.356 | 0.349 | 0.342 | 0.336 |

Table A.4: NDCG results for MSLR-WEB30K Fold 1.

| Model | @1 | @2 | @3 | @4 | @5 | @6 | @7 | @8 | @9 | @10 |
|---|---|---|---|---|---|---|---|---|---|---|
| GBRT | **0.436** | **0.425** | **0.426** | **0.430** | **0.435** | **0.439** | **0.445** | **0.449** | **0.454** | **0.457** |
| RankNet | 0.126 | 0.136 | 0.144 | 0.151 | 0.158 | 0.164 | 0.171 | 0.176 | 0.182 | 0.187 |
| RankBoost | 0.276 | 0.280 | 0.288 | 0.296 | 0.304 | 0.310 | 0.317 | 0.323 | 0.329 | 0.334 |
| AdaRank | 0.215 | 0.230 | 0.242 | 0.253 | 0.263 | 0.270 | 0.278 | 0.285 | 0.292 | 0.298 |
| Coordinate Ascent | 0.424 | 0.401 | 0.397 | 0.397 | 0.399 | 0.401 | 0.404 | 0.408 | 0.412 | 0.416 |
| LambdaMART | 0.355 | 0.358 | 0.364 | 0.369 | 0.376 | 0.381 | 0.387 | 0.392 | 0.398 | 0.402 |
| ListNet | 0.121 | 0.130 | 0.139 | 0.147 | 0.154 | 0.160 | 0.167 | 0.173 | 0.179 | 0.184 |
| Random Forests | 0.373 | 0.366 | 0.371 | 0.375 | 0.382 | 0.389 | 0.394 | 0.401 | 0.407 | 0.412 |

*Author: Sen LEI, Xinzhi HAN*